\documentclass[12pt]{article}

\usepackage{graphicx}
\usepackage{amsfonts}
\newtheorem{proposition}{Proposition}

\newtheorem{definition}[proposition]{Definition}
\newtheorem{example}[proposition]{Example}
\newtheorem{theorem}[proposition]{Theorem}

\def\+{{+\!\!\!+}}

\def\d{\partial}

\def\pmb#1{\setbox0=\hbox{#1}% 
\kern.0em\copy0\kern-\wd0 
\kern-.04em\copy0\kern-\wd0 
\kern.08em\copy0\kern-\wd0 
\kern-.04em\raise.0433em\box0 }         %poor man's bold macro (TexBook) 
\newcommand{\nc}{\newcommand} 
\nc{\beq}{\begin{equation}} 
\nc{\eeq}[1]{\label{#1}\end{equation}} 
\nc{\ber}{\begin{eqnarray}} 
\nc{\eer}[1]{\label{#1}\end{eqnarray}} 
\nc{\pek}[1]{\cite{#1}} 
\nc{\enr}[1]{(\ref{#1})} 
\nc{\kal}[1]{{\cal{#1}}} 
\nc{\dott}{\;\cdot\;} 
\newcommand{\Section}[1]{\section{#1} \setcounter{equation}{0}}

\def\0 {\nonumber}

\begin{document} 
\setcounter{page}{0}
\newcommand{\inv}[1]{{#1}^{-1}} %inverse 
\renewcommand{\theequation}{\thesection.\arabic{equation}} 
\newcommand{\be}{\begin{equation}} 
\newcommand{\ee}{\end{equation}} 
\newcommand{\bea}{\begin{eqnarray}} 
\newcommand{\eea}{\end{eqnarray}} 
\newcommand{\re}[1]{(\ref{#1})} 
\newcommand{\qv}{\quad ,} 
\newcommand{\qp}{\quad .} 
\begin{titlepage} 
%\title{} 
\begin{center} 

\hfill SISSA 51/2005/EP\\                          
\hfill   hep-th/0507051\\

\vskip .3in \noindent 

%\vskip .1in 

{\Large \bf{From current algebras for p-branes\\
to topological M-theory}} \\

\vskip .2in 

{\bf Giulio Bonelli$^{a}$}\footnote{e-mail address: bonelli@sissa.it}
and {\bf Maxim Zabzine$^{b}$}\footnote{e-mail address: m.zabzine@qmul.ac.uk}

\vskip .05in 
$^a${\em\small International School of Advanced Studies (SISSA) and INFN, Sezione di Trieste\\
 via Beirut 2-4, 34014 Trieste, Italy} 
\vskip .05in 
$^b${\em School of Mathematical Sciences, Queen Mary, University of London, \\
Mile End Road, London, E1 4NS, UK}
\vskip .5in 
\end{center} 
\begin{center} {\bf ABSTRACT }  
\end{center} 
\begin{quotation}\noindent  
 In this note we generalize a result by Alekseev and Strobl for the case of $p$-branes.
  We show that there is a relation between anomalous free current algebras and 
   "isotropic" involutive subbundles of $T\oplus \wedge^p T^*$ with the Vinogradov
   bracket, that is a generalization of the Courant bracket.  As an application of this construction
     we go through some interesting examples: topological strings on symplectic manifolds, topological 
     membrane on $G_2$-manifolds and topological 3-brane on $Spin(7)$ manifolds.
    We show that 
   these peculiar topological theories are related to the physical (i.e., Nambu-Goto)
    brane theories in a specific 
   way.  These topological brane theories are proposed as microscopic description of 
    topological 
    M/F-theories.
 \end{quotation} 
\vfill 
\eject 

%\addtocontents{toc}
%\tableofcontents

\end{titlepage}

\section{Introduction}

 Recently Hitchin proposed to consider  the generalized geometry where the tangent bundle
 $TM$ is replaced by the tangent plus cotangent bundle $TM \oplus T^*M$. In different 
 context and by different authors it has been pointed out that there is
 string theory origin of the  generalized geometry based on $TM \oplus T^*M$. 
  Indeed many concepts of generalized geometry have their string theory counterpart.
 Insprired by this relation we would like to make one step further and ask 
 about possible relevant geometric concepts for $p$-brane theories.  In this note 
 we propose that for $p$-brane theories the relevant geometry is based on $TM \oplus \wedge^p T^*M$
 bundle. 

 The paper consists of two results. First of all we generalize Alekseev-Strobl observation
  \cite{Alekseev:2004np} to the case of generic $p$-brane theory. 
 Namely we associte to anomaly free algebra of 
 $p$-brane currents an ``isotropic'' involutive subbundle $L$ of $T\oplus \wedge^p T^*$. 
 This algebra can be regarded as an algebra of first class constraints for some gauge theory.
 In particular we consider a few interesting examples of such gauge theories, namely
 topological $p$-brane theories. We study the compatibility condition between $L$
 and Riemannian geometry and show that it  singles out a
   very interesting subclass of topological 
 $p$-brane theories on special class of manifolds. 
 These examples complement the recent discussion of topological M-theory 
 \cite{Gerasimov:2004yx, Dijkgraaf:2004te, Grassi:2004xr,  Nekrasov:2004vv, Smolin:2005gu}  and topological F-theory  \cite{Anguelova:2004wy}, 
 however at microscopic level. This is our second result.

 The structure of the paper is as follows. In Section \ref{ham} we describe the phase space for 
 $p$-brane theory which is a simple generalization of the cotangent bundle of loop  spaces.
 In Section \ref{dirac} we associate  currents to the sections of $T \oplus \wedge^p T^*$
 and calculate the Poisson bracket between them. The calculation gives rise
 to the Vinogradov bracket on $T \oplus \wedge^p T^*$ (the direct generalization of 
 Courant bracket on $T \oplus T^*$) and a specific anomalous term. The anomaly free subalgebras
 of the currents can be associated with ``isotropic'' involutive subbundles of $T \oplus \wedge^p T^*$. 
 We discuss the examples of such subbundles and show that 
 the anomaly free subalgebras of currents can be interpreted as first class constraints of 
 some gauge theory. In Section \ref{vector} we consider the class of  topological 
 $p$-brane theories which are related to the Nambu-Goto $p$-branes in a specific way.
   Actually we obtain the topological strings on symplectic and K\"ahler manifolds, 
 topological membranes on $G_2$-manifolds and topological $3$-branes on 
 $Spin(7)$-manifolds. 
 Section \ref{open} presents some comments on the open $p$-brane theory. In particular we discuss
 the allowed boundary conditions which preserve the relevant symmetries. 
 In Section \ref{end} we summarize and collect some general comments for the future research. 

\section{Hamiltonian formalism for $p$-branes}
\label{ham}

The phase space  of closed  strings on a manifold $M$
 can be identified with the cotangent bundle $T^*LM$
 of the loop space $LM = \{X: S^1 \rightarrow M \}$. 
 Below we present a straightforward generalization  of this construction
  to the case of generic closed $p$-brane theory. 

 Following the logic above 
 for the $p$-brane world-volume $\Sigma_{p+1}= \Sigma_p \times \mathbb R$ the phase space can be identified with the 
 cotangent bundle $T^* \Sigma_p M$ of the space of maps,  $\Sigma_p M = \{ X: \Sigma_p \rightarrow M\}$. Using 
 local coordinates $X^\mu(\sigma)$ and their conjugate momenta $p_\mu(\sigma)$ the standard symplectic 
 form on $T^*\Sigma_p M$ is given by
\beq
 \omega = \int\limits_{\Sigma_p}d^p\sigma\,\, \delta X^\mu \wedge \delta p_\mu,
\eeq{defsinsymp}
 where $\delta$ is de Rham differential on $T^*\Sigma_p M$. The canonical dimensions of the fields
 should be chosen such that $\omega$ is dimensionless. Namely we choose\footnote{We work in units 
  where $p$-brane tension $T_p$ is  equal to one. For details see Appendix B.} $dim[X^\mu]=0$, $dim[\sigma]=1$
 $dim[\d]=-1$ and $dim[p_\mu] = -p$.
The symplectic form (\ref{defsinsymp})
 can be twisted by a closed $(p+2)$-form $H$, $H \in \Omega^{p+2}(M)$, $dH=0$, as follows
\beq
 \omega = \int\limits_{\Sigma_p}d^p\sigma\,\,\left( \delta X^\mu \wedge \delta p_\mu + \frac{1}{2}
 H_{\mu_1\mu_2\mu_3 ...\mu_{p+2}} \delta X^{\mu_1} \wedge \delta X^{\mu_2} \epsilon^{\alpha_1...\alpha_p}
 \d_{\alpha_1} X^{\mu_3}... \d_{\alpha_p} X^{\mu_{p+2}}\right ),
\eeq{twistedsymp}
 where $\epsilon^{\alpha_1 ...\alpha_p}$ is completely antisymmetric tensor on $\Sigma_p$. 
 The symplectic form (\ref{twistedsymp}) implies the Poisson brackets
\beq
 \{ X^\mu(\sigma), X^\nu(\sigma')\} = 0,\,\,\,\,\,\,\,\,\,\,\,\,\,\,\,\,
 \{ X^\mu(\sigma), p_\nu(\sigma') \} = \delta^\mu_\nu \delta(\sigma - \sigma'),
\eeq{poisbra1}
\beq
 \{ p_\mu(\sigma), p_\nu(\sigma') \} = - H_{\mu\nu\rho_1...\rho_p} \epsilon^{\alpha_1...\alpha_p}
 \d_{\alpha_1} X^{\rho_1} ... \d_{\alpha_p} X^{\rho_p} \delta(\sigma - \sigma').
\eeq{poisbra2}
 For the symplectic structure (\ref{twistedsymp}) the transformation 
\beq
 X^\mu\,\,\rightarrow\,\,X^\mu,\,\,\,\,\,\,\,\,\,\,\,\,\,\,\,\,
 p_\mu\,\,\rightarrow\,\,p_\mu + b_{\mu\nu_1...\nu_p} \epsilon^{\alpha_1 ...\alpha_p} \d_{\alpha_1} X^{\nu_1}
 ...\d_{\alpha_p} X^{\nu_p} 
\eeq{canonicaltrsam}
 is canonical if  $b \in \Omega^{p+1}(M)$, $db = 0$.
 There are also canonical transformations which correspond to $Diff(M)$ when $X$ transforms as
 a coordinate and $p$ as a section of cotangent bundle $T^*M$. Indeed the group of {\it local} canonical transformations
 for $T^*\Sigma_p M$ is a semidirect product of $Diff(M)$ and $\Omega^{p+1}_{closed}(M)$ in analogy 
 with the loop space case \cite{Zabzine:2005qf}.

 Finally we conclude the discussion of Hamiltonian formalism for $p$-brane theory with the following 
 comment. Typically the symplectic form (\ref{twistedsymp}) arises from the action
\beq
 S(\gamma) = \int\limits_{\gamma} \,\,(\theta  - h),
\eeq{ACTSJAPOE} 
 where $\theta$ is a Liouville form $\omega = \delta \theta$, $h$ is a Hamiltonian and 
 $\gamma$ is a path in $T^*\Sigma_pM$.  In order the exponential 
 of this action, $e^{iS(\gamma)}$ to be well-defined we have to impose the intergrality condition 
 on $H$. Namely we have to require that $[H] \in H^{p+2}(M, \mathbb{Z})$.

\section{Current algebra and generalized Dirac structure}
\label{dirac}

In this section we consider the generalization of the idea proposed in \cite{Alekseev:2004np}, where
  the authors established the relation between 2D 
 anomaly free current algebras and Dirac structures.   

 Let us consider the currents which are linear in momentum $p_\mu$.  If we assume
  that the currents do not depend on any 
 dimensionful parameter or world-volume metric then
    the most general form is given by 
\beq
J_{\epsilon} (v+ \omega) = \int\limits_{\Sigma_p} d^p\sigma\,\, \epsilon \left (v^\mu(X) p_\mu
 + \omega_{{\mu_1}...{\mu_p}} (X)\epsilon^{\alpha_1...\alpha_p} \d_{\alpha_1} X^{\nu_1}
 ... \d_{\alpha_p} X^{\nu_p} \right ),
\eeq{defcurrent}
 where $v+\omega$ is a section of $T \oplus \wedge^p T^*$
   and $\epsilon \in C^\infty(\Sigma_p)$ is a test function. 
 Using the symplectic structure (\ref{defsinsymp}) we calculate the Poisson bracket of two 
 currents associated to $(v+\omega), (\lambda+s)  \in C^\infty(T \oplus \wedge^p T^*)$, 
$$ \{ J_{\epsilon_1} (v+\omega), J_{\epsilon_2}(\lambda + s)\}
  = - J_{\epsilon_1\epsilon_2} ([[v+\omega, \lambda +s]])  - $$
  \beq
  -p\int\limits_{\Sigma_p} d^p\sigma\, (\d_{\alpha_1}
\epsilon_1)\epsilon_2 (i_v s + i_\lambda\omega)_{\nu_2...\nu_p} \epsilon^{\alpha_1\alpha_2...\alpha_p}
\d_{\alpha_2} X^{\nu_2} ... \d_{\alpha_p} X^{\nu_p},
\eeq{poisonal1}
 where the bracket $[[\,\,,\,\,]]$ is defined as follows
\beq
 [[v + \omega, \lambda + s]] = [v,\lambda ] + {\cal L}_v s - {\cal L}_\lambda \omega +
  d(i_\lambda \omega).
\eeq{definnowskew}
 In (\ref{definnowskew}) $[\,\,,\,\,]$ is the standard Lie bracket on $TM$ and ${\cal L}$ is 
 a Lie derivative.  
 Alternatively the result (\ref{poisonal1}) can be rewritten as
$$ \{ J_{\epsilon_1} (v+\omega), J_{\epsilon_2}(\lambda + s)\}
  = - J_{\epsilon_1\epsilon_2} ([v+\omega, \lambda +s]_c)  + $$
  \beq
  +\frac{p}{2} \int\limits_{\Sigma_p} d^p\sigma\, (\epsilon_1 \d_{\alpha_1}
\epsilon_2 -  \epsilon_2 \d_{\alpha_1}\epsilon_1)
(i_v s + i_\lambda\omega)_{\nu_2...\nu_p} \epsilon^{\alpha_1\alpha_2...\alpha_p}
\d_{\alpha_2} X^{\nu_2} ... \d_{\alpha_p} X^{\nu_p},
\eeq{poisonal2}
where the bracket $[\,\,,\,\,]_c$ is given by
\beq
 [ v + \omega, \lambda + s]_c = [v,\lambda] + {\cal L}_v s - {\cal L}_\lambda \omega
  - \frac{1}{2} d( i_v s - i_\lambda \omega).
\eeq{deskewsym}
 The bracket $[\,\,,\,\,]_c$ is just antisymmetrization of the bracket $[[\,\,,\,\,]]$. 

 The bracket $[[\,\,,\,\,]]$ is an example of derived bracket (see \cite{kosmann} for a review)
 and its antisymmetrization $[\,\,,\,\,]_c$ is called Vinogradov bracket. 
  One interesting feature  is that the bracket $[\,\,,\,\,]_c$ has non-trivial automorphisms defined
   by forms \cite{hitchin1}. Let $b \in \Omega^{p+1}(M)$ be a closed $(p+1)$-form which defines
    the vector bundle automorphism $e^b$ of $T\oplus \wedge^p T^*$
    \beq
     e^b(v +\omega) \equiv v + \omega + i_v b.
    \eeq{definauto}
    Then the bracket $[\,\,,\,\,]_c$ satisfies 
    \beq
     e^b\left ([v +\omega, \lambda+s]_c\right )=
     [e^b(v+\omega), e^b(\lambda+s)]_c .
    \eeq{actionautobr}
 This non-trivial automorphism of $[\,\,,\,\,]_c$ corresponds to the canonical transformation
  (\ref{canonicaltrsam}) at the level of Poisson bracket of currents (\ref{poisonal2}).
 If we are interested in the situation 
 when anomalous term is absent in (\ref{poisonal2}) and the currents form a closed algebra
 then we should require the following.   Let label the currents by sections of a subbundle
  $L \subset T \oplus \wedge^p T^*$.
 In (\ref{poisonal2}) the anomalous term is absent if for  any $(v + \omega), (\lambda+s) \in  C^\infty(L)$
\beq
 \frac{1}{2} (i_v s + i_\lambda \omega) \equiv 
 \langle v + \omega, \lambda + s \rangle = 0,
\eeq{isotroap}
 where  $\langle\,\,,\,\,\rangle$ is ``pairing'' between two sections of $T\oplus \wedge^p T^*$
 which is a map $(T \oplus \wedge^p T^*) \times  (T \oplus \wedge^p T^*)\,\,\rightarrow \wedge^{p-1} T^*$
 where $\wedge^0 T^*$ is understood as $\mathbb{R}$. The bundle automorphism (\ref{definauto})
 preserves this ``pairing''.
We call isotropic  any subbundle $L$ which satisfies (\ref{isotroap}). 
 Moreover if we require that our currents form a closed subalgebra  then we
 have to  impose that for  any two sections  $(v + \omega), (\lambda+s) \in  C^\infty(L)$ 
 the section $[v+ \omega, \lambda +s]_c \in C^{\infty}(L)$, 
 i.e. the subbundle $L$ is involutive.  
 Indeed the bracket $[\,\,,\,\,]_c$
 restricted to involutive isotropic subbundle of $T \oplus \wedge^p T^*$ is a Lie bracket\footnote{
  Indeed $L$ has a structure of the Lie algebroid with the anchor being a natural projection to $TM$.}.
 Since we could not find the proof of this statement in the literature we present the proof
 in Appendix A as well as other relevant properties of the brackets. The proof is a direct 
 generalization of the proof for $T \oplus T^*$. 
 Thus isotropic involutive subbundle $L$, as defined above, 
 corresponds to anomaly free algebra of currents
\beq
\{ J_{\epsilon_1} (v+\omega), J_{\epsilon_2}(\lambda + s)\}
  = - J_{\epsilon_1\epsilon_2} ([v+\omega, \lambda +s]_c|_L). 
\eeq{anomalfreals}
 For the case $p=1$ if $L$ is also maximally isotropic then it is called Dirac structure. 
 In the general situation $ p \geq 2$ it is tempting to define a generalized Dirac 
 structure as a maximally isotropic involutive subbundle of $T \oplus \wedge^p T^*$. 
 Although we have to admit that the notion of maximality of isotropic condition 
 (\ref{isotroap}) is not very natural, however see some comments in Appendix.  
 For different definitions of generalization of Dirac structure for $T \oplus \wedge^p T^*$
 (also called the Dirac-Nambu structure) see \cite{hagiwara} and \cite{wade}. 

 The algebra of currents (\ref{anomalfreals}) corresponding to involutive isotropic 
 subbundle $L$ can be regarded as an algebra of first class constraints for some 
 gauge theory. In next Section we will give a few examples of such theories, namely 
  topological $p$-branes.  

 Let us present some examples of isotropic involutive subbundles of $T\oplus \wedge^p T^*$. 
\begin{example}
 Let us fix a $(p+1)$-form, $\phi\in \Omega^{p+1}(M)$ and  consider the 
 subbundle $L = \{ v + i_v \phi, v \in T\} \subset T \oplus \wedge^p T^*$ which is obviously 
 isotropic
$$  \langle v + i_v \phi, \lambda + i_\lambda \phi \rangle =
  \frac{1}{2} (i_v i_\lambda \phi + i_\lambda i_v \phi) = 0.$$
 Next calculate the bracket between two sections
\beq
 [v + i_v \phi, \lambda + i_\lambda \phi]_c = [v, \lambda] + i_{[v,\lambda]} \phi +
 i_\lambda i_v d \phi,
\eeq{calculasprbap}
 where we used the property $[{\cal L}_v, i_\lambda ] = i_{[v,\lambda]}$. The subbundle is 
 involutive if the last term vanishes in (\ref{calculasprbap}), i.e. $d\phi =0$. In other words
 $T$ is involutive isotropic subbundle and $L = e^\phi (T)$, where $e^\phi$ is the bundle automorphism 
 defined in (\ref{definauto}) for closed $(p+1)$-form.
\end{example}
 The next example is related to the complexification of the bundle 
 $(T \oplus \wedge^p T^*) \otimes \mathbb{C}$. 
\begin{example}
 On complex manifold we can consider the subbundle $L = T_{(1,0)} \oplus (\wedge^p T^*)_{(0,p)}$ 
 of  $(T \oplus \wedge^p T^*) \otimes \mathbb{C}$. The sections of $L$ are holomorphic vector fields
 and antiholomorphic forms (i.e., elements of $\Omega^{(0,p)}(M)$). The subbundle $L$ is
 obviously isotropic and the bracket of two sections of $L$ is
$$ [v + \omega, \lambda +s]_c = [v,\lambda] + i_v \d s - i_\lambda \d\omega $$
 which is clearly a section of $T_{(1,0)} \oplus (\wedge^p T^*)_{(0,p)}$. Thus $L$ is an isotropic 
 involutive subbundle.  
\end{example}
 It is not hard to produce  other examples of involutive isotropic subbundles of $T \oplus \wedge^p T^*$,
 for example based on foliated geometry. In addition we can apply any closed 
  $(p+1)$-form $b$ which defines automorphism (\ref{definauto}) to
   an isotropic involutive subbundle $L$
   to obtain another isotropic involutive subbundle $e^b(L)$. 

 So far we calculated the Poisson brackets using (\ref{defsinsymp}) as  symplectic structure.
 More generally we can calculate the Poisson brackets (\ref{poisonal2})
  using the twisted symplectic structure (\ref{twistedsymp}) with 
 $H \in \Omega^{p+2}(M)$, $dH=0$. In this case the bracket $[\,\,,\,\,]_c$ in 
 (\ref{poisonal2}) gets replaced by its twisted version
\beq
 [v + \omega, \lambda + s]_H = [v + \omega, \lambda + s]_c + i_v i_\lambda H.
\eeq{twistedbr}
 All considerations above can be generalized to this case. Thus in particular 
   Example 1 gives rise to 
 isotropic involutive (with respect to $[\,\,,\,\,]_H$) subbundle if $d\phi = H$. 

 Finally let us note that the currents (\ref{defcurrent}) behave nicely under the diffeomorphisms of 
 $\Sigma_p$. Introduce the generator of  of $Diff(\Sigma_p)$
 $${\cal H}_\alpha [N^\alpha]=\int\limits_{\Sigma_p}d^p\sigma\,\,
  N^\alpha \partial_\alpha X^\mu p_\mu,$$
 where $N^\alpha$ is a text function. The Poisson bracket between generator of 
 $Diff(\Sigma_p)$ and the current (\ref{defcurrent}) is
$$\{ {\cal H}_\alpha[N^\alpha] ,J_\epsilon (v+\omega)\}=
J_{N^\alpha\partial_\alpha\epsilon} (v+\omega),$$
 where we assume (\ref{twistedsymp}) as symplectic structure.

\section{Vector cross product and topological branes}
\label{vector}

In this Section we use the construction of involutive isotropic subbundle $L$
 given in Example 1 from previous Section. For this subbundle we can construct 
 the anomaly free subalgebra of currents (\ref{anomalfreals}).   We  interpret
 these currents
 as first class constraints for a topological p-brane theory.
 We impose a specific compatibility of $\phi$ with a Riemannian metric $g$ on $M$ which
  leads to a certain relation between topological  and 
  physical p-brane theories. Indeed all  such theories can be classified and there is
  a finite number of them.

 We start by explaining the compatibility condition between the $(p+1)$-form $\phi$ and a Riemannian 
 metric $g$ on $M$. 
  We all are familiar with the usual vector cross product $\times$ of two vectors in $\mathbb{R}^3$, which 
 satisfies 

$\bullet$ $u \times v$ is bilinear and skew symmetric

$\bullet$ $ u \times v \perp u, v$; so $(u\times v) \cdot v=0$  and $(u\times v)\cdot u=0$

$\bullet$ $(u \times v) \cdot (u \times v) = \det \left ( \begin{array}{ll}
                                                  u \cdot u & u \cdot v\\
                                                  v\cdot u & v \cdot v
                                             \end{array} \right )$\\

 The generalization of vector cross product to a Riemannian manifold
  leads to the following definition by Brown and Gray \cite{brown}

\begin{definition}
 On $d$-dimensional Riemannian manifold $M$ with a metric $g$ an $p$-fold vector cross product
 is a smooth bundle map
$$  \chi : \wedge^p TM \rightarrow TM $$
satisfying
$$g (\chi(v_1,...,v_p), v_i)=0,\,\,\,\,\,\,1 \leq i \leq p$$
$$ g(\chi(v_1,...,v_p), \chi(v_1,...,v_p)) = \| v_1 \wedge ... \wedge v_p \|^2$$
 where $\|...\|$ is the induced metric on $\wedge^p TM$.
\end{definition}
 Equivalently the last property can be rewritten in the following form
$$ g(\chi(v_1,...,v_p), \chi(v_1,...,v_p)) = \det ( g(v_i, v_j))= \| v_1 \wedge ... \wedge v_p \|^2. $$
 The first condition in the above definition is equivalent to the following tensor $\phi$
$$ \phi (v_1,...,v_p, v_{p+1}) = g (\chi(v_1,...,v_p), v_{p+1})$$
 being a skew symmetric tensor of degree $p+1$, i.e. $\phi \in \Omega^{p+1}(M)$. 
 Thus in what follows we consider the $(p+1)$-form $\phi$ which defines the $p$-fold vector 
 cross product. 

Cross product on real spaces  were classified by Brown and Gray \cite{brown}. 
 The global vector cross products on manifolds were first studied by Gray \cite{gray}.
 They fall into four categories:

(1) With $p=d-1$ and $\phi$ is the volume form of manifold

(2) When $d$ is even and $p=1$, we can have a one-fold cross product $J: TM \rightarrow TM$. 
 Such a map satisfies $J^2=-1$ and is almost complex structure. The associated $2$-form 
 is the K\"ahler form.

(3) The first of two exceptional cases is a $2$-fold cross product ($p=2$) on a $7$-manifold.
 Such a structure  is called a $G_2$-structure and the associated $3$-form is called 
 a $G_2$-form.

(4) The second exceptional case is $3$-fold cross product ($p=3$) on 
 $8$-manifold. This is called a $Spin(7)$-structure and the associated 
 $4$-form is called $Spin(7)$-form.

 Notice that there are  similarities of this list of real vector cross products 
 with the list of   stable forms \cite{hitchin2}. Namely the cases (2) and  (3)  
  correspond to stability of $\phi$.  The complexified version of 
 the vector cross product which allows to consider Calabi-Yau manifolds, see \cite{complex}.
  However we will not review the complex version of vector cross product. 

 Following the discussion from previous section, in particular Example 1,
  there is a set of topological $p$-brane theories we can associate to a $p$-fold 
  vector cross product characterized by $(p+1)$-form $\phi$.
 Consider a subbundle $L=\{v+ i_v \phi, v \in T\}$ of $T \oplus \wedge^p T^*$. 
 To the sections of $L$  we can associate the following constraints (currents)
\beq
 J_\mu = p_\mu + \phi_{\mu\nu_1...\nu_p} \epsilon^{\alpha_1...\alpha_p} \d_{\alpha_1}X^{\nu_1} 
...\d_{\alpha_p} X^{\nu_p} = 0,
\eeq{curentspa}
 where we work in local basis $\d_\mu$. Alternatively we can rewrite the constraints in 
 coordinate free form
\beq
 i_v J = i_v p + g (\chi (\d_1 X, ..., \d_p X), v)  = 0,
\eeq{constrainq;d}
 where $v$ is a section of $TM$.
 The constraints (\ref{curentspa}) are the first class with respect to the symplectic 
 form (\ref{defsinsymp}) if $d \phi = 0$.  In the twisted case, when one uses (\ref{twistedsymp}), 
  the first class condition leads to $d\phi = H$.
 
 Let us now study the compatibility condition between the topological system (\ref{curentspa})
  and the Nambu-Goto dynamics.
The constraints (\ref{curentspa}) imply the Nambu-Goto 
costraints (see Appendix B) 
\ber
&&{\cal H}_\alpha = p_\mu \d_\alpha X^\mu = 0\label{AAAA123}\\
&&{\cal H} = p_\mu g^{\mu\nu} p_\nu - \det (\d_\alpha X^\mu g_{\mu\nu} \d_\beta X^\nu) =0\label{AAAA1234}
\eer{cosnatei11}
 if and only if $\phi$ corresponds to vector cross product\footnote{Indeed previously 
  the cross vector  product has been discussed in the context of $p$-brane instantons
   for the Nambu-Goto theory \cite{Biran:1987ae, Grabowski:1989cx}.}. Namely
$$ {\cal H}_\alpha = J_\mu \d_\alpha X^\mu =0$$
 and 
$$p_\mu g^{\mu\nu} p_\nu = \| \d_1 X\wedge ... \wedge \d_p X \| = \det (g (\d_\alpha X, \d_\beta X)), $$
 where we have used the second property in the definition of vector cross product. 
 Indeed the Nambu-Goto $p$-brane theory is decribed by $(p+1)$ constraints (\ref{AAAA123}) and (\ref{AAAA1234}), see 
 Appendix B for the details. 

 We constructed TFTs such that their constraint surface $J_\mu = 0$ lies inside
 the constraint surface for the standard $p$-brane theory,
$$ J_\mu = 0 \,\,\,\,\,\,\, \Rightarrow \,\,\,\,\,\,\, {\cal H}_\alpha=0,\,\,\,{\cal H}=0.$$
  Classically it means that the BRST cohomology of topological branes is subspace of 
 the BRST cohomology  of physical brane theory. At quantum level we may speculate
 that the correlators of observables of topological brane theory are related to subsector 
 of physical brane theory, in analogy with the relation between topological strings
  and superstrings. However, at the present level of discussion, we cannot elaborate more 
 on the relation between quantum toopological and physical brane theories. 
 
  There is an alternative point of view on the relation between the topological
 $p$-brane theory and standard $p$-brane theory (i.e., given by Nambu-Goto (NG) action)
  on a manifold with a vector cross product structure. Namely the Nambu-Goto action can 
   be thought of as a deformation of the corresponding topological theory.
The Hamiltonian of Nambu-Goto theory is given by the following expression
$$h_{NG}=\int  d^{p}\sigma\,\, \left (N{\cal H} + N^\alpha {\cal H}_\alpha\right),$$
 where ${\cal H}$, ${\cal H}_\alpha$ are the constraints (\ref{AAAA123})-(\ref{AAAA1234})
  and $N$, $N^\alpha$ are Lagrangian multipliers. Next assume that $\phi$ defines 
   a vector cross product with respect to $g$, so does $-\phi$.  We define the currents
 $$ J_\mu^{\pm} = p_\mu \pm \phi_{\mu\nu_1...\nu_p} \epsilon^{\alpha_1...\alpha_p} \d_{\alpha_1}X^{\nu_1} 
...\d_{\alpha_p} X^{\nu_p}$$ 
 and rewrite the constraints ${\cal H}$, ${\cal H}_\alpha$ as follows
$$
{\cal H}= J^+_{\mu}g^{\mu\nu} J^-_{\nu}
\quad {\rm and}\quad
{\cal H}_\alpha=\partial_\alpha X^{\mu}J^+_\mu,
$$
 where we used the fact that $\phi$ is vector cross product with respect to $g$.
As a further preparatory step, we introduce the auxiliary fields $B_{\pm}^\mu$
and rewrite the NG action as
\beq
S_{NG}=\int dt \,d^p\sigma\,\,\left ( p_\mu \dot X^\mu  - B^\mu_{-} J^-_{\mu}-
 B^\mu_{+} J^+_{\mu}
+\frac{1}{N}B_+^\mu g_{\mu\nu}B_-^\nu
 - \frac{1}{N} N^\alpha
\partial_\alpha X^{\mu}
 g_{\mu\nu} B^\nu_- \right )
\eeq{actskwp29}
Since the fields $B_+^\mu$ enter linearly we can integrate them and arrive at the standard 
 Nambu-Goto action in the phase space form.  Obviously the action (\ref{actskwp29})
  is not unique and there are other equivalent ways to rewrite it. 

 For the topological $p$-brane theory we have two possible (equivalent) Hamiltonians
$$
h^{\pm}=\int d^p\sigma\,\,B^\mu_{\pm} J^\pm_{\mu},
$$
where $B^\mu_{\pm}$ are the Lagrange multipliers.
    In action (\ref{actskwp29}) actually  both  currents $J^{\pm}_\mu$ enter. However
we do not want to introduce two copies of the topological theory and thus one of the two should be fake.
This can be easily obtained by considering the action
\beq
S_{top}=\int dt\,d^p\sigma \left ( p_\mu\dot X^\mu - B^\mu_- J^-_{\mu}-B^\mu_+ J^+_{\mu}
-\chi_\mu B^\mu_+ \right ),
\eeq{topaction1}
where $\chi_\mu$ is the Lagrange multiplier freezing $B^\mu_+$.
 Now combining (\ref{actskwp29}) and (\ref{topaction1}) it is straightforward to write 
  the Nambu-Goto action as follows 
\beq
S_{NG}=S_{top} - \lambda S_{def}
\eeq{fulldefkow}
where
$$
S_{def}=\int dt\,d^p\sigma \left (
\frac{1}{N} N^\alpha\partial_\alpha X^{\mu} g_{\mu\nu} B_-^\mu
+
\eta_\mu\left(g^{\mu\nu}\chi_\nu +\frac{1}{N}B_-^\mu\right)
\right )
$$
where $\eta$ is an additional auxiliary field. 
In (\ref{fulldefkow}) at $\lambda=0$ the theory describes the topological $p$-brane theory.
 If $\lambda$ is non zero  then the action (\ref{fulldefkow})  becomes
$S_{NG}$ upon a rescaling of the Lagrange multipliers $N^\alpha\to\lambda^{-1} N^\alpha$.
This construction (or its versions) exists only if $\phi$ corresponds to a vector cross product structure and 
 $d\phi =0$.

 Thus for the list of vector cross product structures given above there is a corresponding 
 list of topological $p$-brane theories. The first case with $\phi$ given by the volume structure 
 corresponds to the trivial case when the Nambu-Goto action is itself topological since it describes 
 the embedding of $(d-1)$-branes into a $d$-dimensional manifold, for details see \cite{Bengtsson:2000xa}. 

 We would like  to discuss the other three non-trivial cases: topological strings on symplectic manifolds
  (also on generalized K\"ahler manifolds),
 topological membranes on $G_2$-manifolds and topological $3$-branes on $Spin(7)$-manifolds. 
 
\subsection{Topological strings on symplectic manifolds}

Case (2) in the list of real vector cross products corresponds to A-model topological strings.
 $1$-fold cross product $J:TM\rightarrow TM$ corresponds to an almost complex structure\footnote{The vector cross product properties read $(gJ)^t=-gJ$ and $J^tgJ=g$ which imply $J^2=-1$.}, $J^2=-1$. 
 The associated $2$-form $\omega= gJ$ is the K\"ahler form. The constraints corresponding to 
 maximally isotropic subbundle $L=\{v + i_v \omega, v \in T\}$ of $T\oplus T^*$ are 
\beq
 p_\mu + \omega_{\mu\nu} \d X^\nu  =0.
\eeq{sympslcase}
 They are first class constraints if $d\omega =0$ and thus the manifold $M$ is symplectic. 
 Indeed this is nothing but A-model topological string theory. 

 As far as classical B-model is concern we have to introduce another structure on $M$. This would 
  correspond to Example 2 in Section \ref{dirac} with $p=1$. Thus in this case $M$ 
   is a complex manifold with the complex structure $J$ and the constraints are 
   given by
   $$ p_i =0,\,\,\,\,\,\,\,\,\,\,\,\,\,\,\,\,
 \d X^{\bar{i}}=0$$
in complex coordinates.
To accomodate both A- and B-models on the same $M$   
 we have to restrict ourselves to the case of K\"ahler manifold $(J, g, \omega = gJ)$. In this case
   we have the following 
 decomposition into holomorphic (antiholomorphic) subbundles
\beq
 (T \oplus T^*)\otimes \mathbb{C} = T^{(1,0)} \oplus T^{(0,1)} \oplus T^{*(1,0)} \oplus T^{*(0,1)}.
\eeq{decompslal}
 There are two interesting sets of complex Dirac structures, first one is $T^{(1,0)} \oplus T^{*(0,1)}$
 (or complementary $T^{(0,1)}\oplus T^{*(1,0)}$) and second is $T^{(1,0)} \oplus T^{*(1,0)}$
 (or complementary $T^{(0,1)}\oplus T^{*(0,1)}$).  Indeed they corresponds to two different 
  generalized complex structures 
  $${\cal J}_{i} : (T \oplus T^*)\otimes \mathbb{C}
   \rightarrow (T \oplus T^*)\otimes \mathbb{C},\,\,\,\,\,\,\,\,\,\,\,i=1,2$$
   such that ${\cal J}_{i}^2 = -1$ and $\Pi^i_{\pm} = \frac{1}{2}(1 \pm i {\cal J}_i)$  project
    maximally isotropic involutive subbundles of $(T \oplus T^*)\otimes \mathbb{C} $
     (for more details see \cite{gualtieri}). In the case of K\"ahler manifolds the corresponding 
      generalized complex structures are 
\beq   
   {\cal J}_1 = \left ( \begin{array}{ll}
      J & \,\,\,\,\,0 \\
      0 & - J^t \end{array} \right ),\,\,\,\,\,\,\,\,\,\,\,\,\,\,\,\,\,\,\,\,\,\,\,\,
    {\cal J}_2 = \left ( \begin{array}{ll}
      0 & - \omega^{-1} \\
      \omega & \,\,\,\,\, 0\end{array} \right )    ,
 \eeq{generlcomspa}
 which commute and give rise to the following positive metric on $T \oplus T^*$ 
$$  {\cal G} = - {\cal J}_1 {\cal J}_2 = \left ( \begin{array} {ll}
 0 & g^{-1}\\
 g & \,\,0 \end{array} \right ). $$ 
 Introducing 
 $$ \Lambda = \left ( \begin{array}{l}
                                   i \d X\\
                                    \,\,\,p\end{array}\right )$$
 as a section of pull-back of tangent and cotangent bundle, $X^*((T \oplus T^*)\otimes \mathbb{C} )$
  we have four topological string theories given by the set of first class constraints
  \beq
   \Pi_\pm^i \Lambda =0.
  \eeq{tioakdo}
  Indeed there are only two distinct theories.
  For the case $T^{(1,0)} \oplus T^{*(0,1)}$ we have $\Pi_-^2\Lambda=0$, i.e.
\beq
 p_i - i g_{i\bar{j}} \d X^{\bar{j}}=0,\,\,\,\,\,\,\,\,\,\,\,\,\,\,
 p_{\bar{i}} + ig_{\bar{i}j} \d X^{j}=0
\eeq{consatajspw}
 which is A-model topological strings. 
 For the other case $T^{(1,0)} \oplus T^{*(1,0)}$  the constraints are $\Pi^1_-\Lambda=0$, i.e.
\beq
 p_i =0,\,\,\,\,\,\,\,\,\,\,\,\,\,\,\,\,
 \d X^{\bar{i}}=0
\eeq{bmodelahdpa}
 corresponding to B-model topological strings\footnote{Using the relation $p_\mu=g_{\mu\nu}\dot{X}$
 in (\ref{consatajspw}) and (\ref{bmodelahdpa}) one can recoginize the holomorphic map and constant 
 map conditions over which A- and B-model path integrals are localized correspondently.}.
   Obviously both A- and B-models constraints imply 
 the physical string constraints, ${\cal H}_1 = p_\mu \d X^\mu=0$ and ${\cal H}= p_\mu g^{\mu\nu}
 p_\nu - \d X^\mu g_{\mu\nu}\d X^\nu=0$. Using the natural pairing $\langle \,\,,\,\,\rangle$
  on $T \oplus T^*$ (see (\ref{pairingtwo}) for $p=1$)
     we can rewrite the string constraints as follows
  \beq
  -i {\cal H}_1 = \langle \Lambda, \Lambda \rangle =0,\,\,\,\,\,\,\,\,\,\,\,\,\,\,\,
   2 {\cal H} = \langle \Lambda, {\cal G} \Lambda \rangle =0.
   \eeq{TTcovstriwpp}
  Since we have formulated everything in $T\oplus T^*$ covariant language it is not 
   hard to generalize above discussion to the case (twisted)
     generalized K\"ahler manifolds as defined in \cite{gualtieri}. 
  The generalized K\"ahler structure is given by two generalized complex structures
   ${\cal J}_1$ and ${\cal J}_2$ which commute and ${\cal G} = -{\cal J}_1 {\cal J}_2$
    defines the positive metric on $T\oplus T^*$.

\subsection{Topological membrane on $G_2$ manifolds}

The first exceptional case, namely (3) in the list of real vector cross product structures, corresponds to
 $M$ being oriented $7$-manifold with a global $2$-fold cross product structure ($p=2$). 
 This cross product is defined by  Riemannian metric $g$ and $3$-form $\Phi$ which
  gives rise to a $G_2$-structure on the manifold\footnote{In this case
   the metric $g$ can be expressed in terms of $\Phi$, \cite{karigiannis}.}. The topological membrane 
  theory on $G_2$-manifold is defined by the following first class constraints in $T^*\Sigma_2M$
\beq
 p_\mu + \Phi_{\mu\nu\rho} \epsilon^{\alpha\beta} \d_\alpha X^\nu \d_\beta X^\rho =0.
\eeq{constraintsp2-}
 The algebraic properties of $\Phi$ are such that the constraints (\ref{constraintsp2-})
  imply the membrane constraints (\ref{AAAA123})-(\ref{AAAA1234}).
 $d\Phi=0$ is equivalent to the fact that (\ref{constraintsp2-}) are first class constraints with 
  respect to the symplectic structure (\ref{defsinsymp}). We put forward this as
  the Hamiltonian description 
   of recently proposed topological M-theory \cite{Gerasimov:2004yx, Dijkgraaf:2004te, Grassi:2004xr,  Nekrasov:2004vv, Smolin:2005gu}
    at microscopic level.

 Suppose that $G_2$-manifold $M_7$ is of the form
 $M_7 = M_6 \times S^1$, 
  where $M_6$ is a six-dimensional manifold with $SU(3)$ structure. Let 
  $X^7$ be a coordinate along $S^1$ then $\Phi$ can be written as 
\beq
 \Phi = \omega \wedge dX^7 + \rho,
\eeq{defoalwif}
 where $\omega$ is the K\"ahler  $2$-form and $\rho$ is the $3$-form which defines
 the almost complex structure on $6$-manifold. If $\omega$ and $\rho$ do not depend on $X^7$
 then $d\Phi =0$ implies that $d\omega =0$ and $d\rho=0$ on $M_6$. 
  Membranes on such $M_7$ can be reduced either to strings on $M_6$ or to membranes 
   on $M_6$ depending on the orientation with respect to $S^1$. If the brane is wrapped 
    along $S^1$ then we can make a partial gauge fixing 
 $X^7 = \sigma_2/L$ with $L$ being the size of $S^1$. Then the constraint
 (\ref{constraintsp2-}) becomes 
\beq
 L p_n + 2 L \rho_{nml} \d_1 X^m \d_2 X^l + 2 \omega_{nm} \d_1 X^m = 0,\,\,\,\,\,\,\,\,\,
 p_7 + 2 \omega_{nm} \d_1 X^n \d_2 X^m =0 ,
\eeq{newcoankfpe}
 where $\mu =(n,7)$.
 If we want to reinterpret this as a constraint in $M_6$ we have to redefine the momenta\footnote{See
 Section \ref{ham} and Appendix for  our 
 conventions on the dimensionality of fields.}
 $p_n |_{M_6}\equiv L p_n$ and restrict our attention only to $\sigma_2$ inedpendent 
 configurations (e.g., by requiring $\d_2 X^n=0$). Assuming this we arrive to 
 the constraint
\beq
 p_n +2 \omega_{nm} \d_1 X^m =0
\eeq{conskahfdopap}
 which is A-model on $M_6$. Another possibility corresponds to the case when original 
  membrane does not have excitations along $X^7$, e.g. $X^7$ chosen to be a constant.
  Then in this case the theory on $M_6$ is membrane theory\footnote{Using the notion 
   of complex vector cross product we can show that a complex version of the constraints
    (\ref{membajdel}) implies the membrane Nambu-Goto constraints.},
  \beq
   p_n + \rho_{nml} \d_1 X^m \d_2 X^l =0.
  \eeq{membajdel}
   Since this theory depends on complex moduli it is tempting to call it B-model.
    Although perturbative B-model is typically defined as a topological string theory
      there should be a dual 
     formulation in terms of membrane theory. Indeed this option is very natural from 
     geometrical point of view due to the moduli dependence.

\subsection{Topological 3-brane on $Spin(7)$ manifolds}

The last case in the list of real vector cross products to
 an oriented $8$-manifold $M$ with a global cross product structure with $p=3$. This cross
 product gives rise to an associated Riemannian metric $g$ and $4$-form $\Psi$. Indeed
 $\Psi$ is self-dual form $*\Psi = \Psi$, which is called sometime Cayley form and defines 
  $Spin(7)$-structure on $M$.
 The theory is described by the following first class constraints in $T^*\Sigma_3M$
\beq
 p_{\mu} + \Psi_{\mu\nu\rho\sigma} \epsilon^{\alpha\beta\gamma} \d_\alpha X^\nu 
 \d_\beta X^\rho \d_\gamma X^\sigma =0.
\eeq{constaruw2910} 
 The algebraic properties of $\Psi$ would follow from the requirement that above constraints
 imply the $3$-brane constraints (\ref{AAAA123})-(\ref{AAAA1234}).
 The closure of $\Psi$ is equivalent to the constraints (\ref{constaruw2910}) being first class 
 with respect to the symplectic structure (\ref{defsinsymp}).  We propose that 
 this topological $3$-brane theory is microscopic description of topological F-theory recently 
 discussed in \cite{Anguelova:2004wy}.

 Let us study two possible reductions of $3$-brane topological theory on $Spin(7)$-manifold
  down to $G_2$- and $SU(3)$-manifolds.
 As a first case consider $Spin(7)$-manifold of the form $M_8 = M_7 \times S^1$ with
\beq
 \Psi = dX^8 \wedge \Phi + * \Phi
\eeq{reducts7manf}
 where $\Phi$ is $G_2$-structure on $M_7$ independent on $X^8$.
  As result $d\Psi=0$ implies $d\Phi = 0$ and $d*\Phi=0$.
 In analogy with the reduction we discussed in previous subsection 
 a reduction of topological $3$-brane theory on $M_8$ gives a topological 
 membrane theory (with $\Phi$ in constraint) theory and topological $3$-brane theory
  (with $*\Phi$ in constraint) on $M_7$.   However topological $3$-brane theory 
   cannot be related to $3$-brane Nambu-Goto theory in a way described previously.

  Following  \cite{Anguelova:2004wy} we can consider $Spin(7)$-manifold
$M_8 = M_6 \times T^2$ where $M_6$ is $SU(3)$-manifold. Assuming that 
 $(X^7, X^8)$ are coordinates along $T^2$ the Cayley form is given by
\beq
 \Psi = dX^7 \wedge \rho  - dX^8 \wedge \hat{\rho} + dX^7\wedge dX^8 \wedge \omega +
  \frac{1}{2} \omega \wedge \omega,
\eeq{refgdjaow}
 where $(\rho, \omega)$ defines  $SU(3)$-structure on $M_6$, such that 
  $\Omega =\rho +i\hat{\rho}$. We can reduce the topological $3$-brane theory given 
   by (\ref{constaruw2910}) down to $M_6$. We get a family of topological theories: 
    topological strings ($\omega$), topological $3$-branes ($\omega \wedge \omega$) and 
     two topological membranes (for $\rho$ and $-\hat{\rho}$). Since on $M_8$
     topological $3$-brane theory is self-dual ($\Psi = *\Psi$), in $M_6$ we get the duality 
      between topological string ($\omega$) and topological $3$-brane ($\omega\wedge\omega$)
       and another duality between topological membrane theories ($\rho$ and $-\hat{\rho}$).
        Indeed two first theories can be interpreted as A-model and membrane theories as
        B-model. This would agree with the expected moduli dependence. 
         Presumably the duality we just discussed is related to proposed S-duality \cite{Nekrasov:2004js}.

\section{Open $p$-branes}
\label{open}

The open string phase space  can be identified with the cotangent 
 bundle $T^* PM$ of the path space $PM= \{X: [0,1] \rightarrow M\, X(0) \in D_0, X(1) \in D_1\}$.
This construction can be generalized to the case of open $p$-branes. 
 Assume for the sake of
 clarity that $\d \Sigma_p$ consists of one component. For such open $p$-brane
 the phase space can be identified with the cotangent bundle $T^* \Sigma_p M_D$ of the space $\Sigma_p M_D =
 \{ X: \Sigma_p \rightarrow M, X(\d \Sigma_p) \subset D\}$ where $D$ is a submanifold of $M$, $i: D \hookrightarrow M$.
  To write 
 down the symplectic structure on $T^*\Sigma_p M_D$ we have to require that there exists $B \in \Omega^{p+1}(D)$ 
 such that $dB = i^* H$. Hence the symplectic structure is given by
$$ \omega = \int\limits_{\Sigma_p}d^p\sigma\,\,
 \left( \delta X^\mu \wedge \delta p_\mu + \frac{1}{2}
 H_{\mu_1\mu_2\mu_3 ...\mu_{p+2}} \delta X^{\mu_1} \wedge \delta X^{\mu_2} \epsilon^{\alpha_1...\alpha_p}
 \d_{\alpha_1} X^{\mu_3}... \d_{\alpha_p} X^{\mu_{p+2}}\right ) -$$
\beq
-  \frac{1}{2} \int\limits_{\d\Sigma_p}d^{p-1}\sigma\,\,
 B_{\mu_1\mu_2\mu_3 ...\mu_{p+1}} \delta X^{\mu_1} \wedge \delta X^{\mu_2} \epsilon^{\alpha_1...\alpha_{p-1}}
 \d_{\alpha_1} X^{\mu_3}... \d_{\alpha_{p-1}} X^{\mu_{p+1}},
\eeq{defsympopen}
 where the boundary contributions are needed in order $\omega$ to be closed, $\delta \omega =0$. 
  If we require the symplectic form (\ref{defsympopen}) to be compatible  with the action (\ref{ACTSJAPOE}) with $\theta$ being 
   a Liouville form for $\omega =\delta \theta$ 
   then, in order to the exponent of this action to be well-defined, we have to 
    impose  $[(H,B)] \in H^{p+2}(M,D, \mathbb{Z})$, where $H^{p+2}(M,D,\mathbb{Z})$ is an integer relative  cohomology group. 
 
 Let us introduce a few useful mathematical notions which are generalizations of the ideas 
 from \cite{gualtieri} used in the context of $T\oplus T^*$.  

\begin{definition}
 Let $M$ be a manifold with a closed $(p+2)$-form $H$. Then the pair $(D,B)$ of a 
  submanifold $i:D \hookrightarrow M$ together with a $(p+1)$-form $B \in \Omega^{p+1}(D)$
   is a generalized submanifold of $(M,H)$  iff $dB= i^*H$.
\end{definition}
 
 A generalized submanifold $(D, B)$ is exactly the data we need to construct 
 the phase space $T^*\Sigma_pM_D$ together with the symplectic structure (\ref{defsympopen}). 

\begin{definition}
 The generalized tangent bundle $\tau_D^B$ of the generalized submanifold $(D,B)$ is 
 $$\tau_D^B =\{ v + \omega \in TD \oplus \wedge^p T^*M|_D: \omega|_D= i_v B\}$$
 isotropic subbundle of $(TM \oplus \wedge^p T^*M)|_D$.
\end{definition}
 If we choose $B=0$ then $\tau_D^0= TD \oplus \wedge^p N^*D$, where 
  $N^*D$ is the conormal subbundle of  the submanifold 
  $D$ (in other word $N^*D= Ann\, TD \subset T^*M$).
 The action of the non-trivial automorphism (\ref{definauto}) of $TM \oplus \wedge^p T^*M$ 
  on generalized submanifolds is given as follows
$$ e^b (D,B) = (D, B+ b)$$

 First consider the simple case when $H=0$ and $B=0$. Introducing the currents (\ref{defcurrent})
 labelled by the section of subbundle $L$ of 
  $TM\oplus \wedge ^p T^*M$ we can calculate their Poisson bracket 
 with respect to the symplectic structure (\ref{defsinsymp}).
 Thus in the case of boundary the calculation (\ref{poisonal2}) is modified
 $$ \{ J_{\epsilon_1} (v+\omega), J_{\epsilon_2}(\lambda + s)\}
  = - J_{\epsilon_1\epsilon_2} ([v+\omega, \lambda +s]_c)  + $$
  $$ +\frac{p}{2} \int\limits_{\Sigma_p} d^p\sigma\, (\epsilon_1 \d_{\alpha_1}
\epsilon_2 -  \epsilon_2 \d_{\alpha_1}\epsilon_1)
(i_v s + i_\lambda\omega)_{\nu_2...\nu_p} \epsilon^{\alpha_1\alpha_2...\alpha_p}
\d_{\alpha_2} X^{\nu_2} ... \d_{\alpha_p} X^{\nu_p} +$$
\beq
+ \frac{1}{2} \int\limits_{\d\Sigma_p} d^{p-1}\sigma\, \epsilon_1 \epsilon_2 
(i_\lambda\omega- i_v s)_{\nu_2...\nu_p} \epsilon^{\alpha_2\alpha_3...\alpha_p}
\d_{\alpha_2} X^{\nu_2} ... \d_{\alpha_p} X^{\nu_p}.
\eeq{curcalcbors}
 As discussed in Section \ref{dirac}
   we have to require that $L$ is an isotropic and involutive subbundle of $TM \oplus \wedge^p T^*M$.
  However now we have to take care of the boundary term in (\ref{curcalcbors}) to the anomaly.
   This can be done by requiring that
   $$ (i_\lambda\omega- i_v s)|_D = 0$$
   for any $(v+\omega), (\lambda+s) \in C^\infty(L)$.  Moreover we have to insure that 
    the action of the currents (i.e., the transformations they generate) do not change the 
     boundary conditions, $X(\d \Sigma_p) \subset D$, i.e. $v$ and $\lambda$ restricted 
      to $D$ should be the sections of $TD$.
      We can fulfill these two conditions together 
    with the isotropy condition of $L$ by the following
   $$ L|_D \,\,\subset\,\, TD \oplus \wedge^p N^*D,$$
  where $L|_D$ is the restriction of subbundle $L$ to the submanifold $D$.
   In the general situation if we allow a generalized submanifold $(D,B)$ then 
    the correct condition is
 \beq
 L|_D \,\,\subset \,\,\tau_D^B,
\eeq{maincondkal} 
i.e. $L|_D$ is a subbundle of the generalized tangent bundle of the generalized 
 submanifold $(D,B)$. 

\section{Conclusions}
\label{end}

Let us first of all summarize what we have been finding in the 
previous sections.
We started by studying specific current algebras for extended objects
requiring the currents to be linear in the momenta,
do not involve any world-volume metric
and
do not contain any dimensionfull parameter.
The current algebras where shown to close under the (twisted or 
untwisted) Poisson bracket if their structure is parametrized by an "isotropic"
involutive subbundle of $T\oplus\wedge^p T^*$.
We may interpreted then these currents as first class constraints for 
topological p-branes theories.

In order to link with the usual Nambu-Goto theory, we required the gauge 
constraints of the topological theory to imply the ones defining the
NG theory itself. Equivalently, we required the topological brane theory
to be a topological truncation of the NG one.
We have shown that the above requirements, namely the algebra closure
and the deformability to the NG theory, 
correspond to the existence of
a {\it real} cross vector product on the manifold 
on which the p-brane theory is formulated.
This mathematical condition reveals to be quite restrictive leaving 
with few well defined cases.
These, and the induced p-brane topological theories, were listed
and analised.
 One of them was 
the A-model topological string in six dimensions, which we reconstruct
in detail.
Through an alternative scheme, we reconstructed the B-model in its usual
formulation too.
In seven dimensions we encountered membrane theory on $G_2$ manifolds
which upon reduction to six dimensions gave the A-model and a novel
membrane theory {\it naturally} coupled to the complex moduli of the six 
manifold.
Analogous phenomena appeared in the last case of 3-branes on eight 
dimensional manifolds admitting a $Spin(7)$ structure.

The reduction of topological F-theory from $Spin(7)$-manifold down 
 to $SU(3)$-manifold produces a whole set of topological brane theories. 
 Some of them are related to Nambu-Goto theories in the way described above.
 One is the topological membrane theory which should be a 
version of the B-model since it couples naturally to the complex moduli. 
 This should be regarded as the nonperturbative completition of 
the A-model. The whole picture requires further study especially at the quantum level.
 We believe that the present reduction can be generalized to BV set-up\footnote{For some
  discussion of BV formalism applied to open topological membrane see \cite{Park:2000au, Hofman:2002jz}.}
  and we hope to come back to this issue in future.

\noindent{\bf Acknowledgement}:  We are grateful to Nigel Hitchin, Simon Lyakhovich,
 Alessandro Tanzini 
 and Pierre Vanhove for discussions. 
The research of G.B. is supported by
the Marie Curie European Reintegration Grant MERG-CT-2004-516466,
the European Commission RTN Program MRTN-CT-2004-005104 and by MIUR.
M.Z. thanks SISSA (Trieste) where part of work was carried out. 
The research of M.Z. was supported by EU-grant
MEIF-CT-2004-500267. 

\noindent{\bf Note added in Proof}:
 After we have finished this work we became aware of two interesting works.
In  \cite{Figueroa-O'Farrill:2005uz}  the authors discuss the gauging of sigma model
 with boundary. Motivated by their example they argue that the notion of isotropic
  subbundle  (\ref{isotroap})
  can be extended to  
  $$\frac{1}{2} (i_v s + i_\lambda \omega) \equiv 
 \langle v + \omega, \lambda + s \rangle = d q,$$
 where $q \in \Omega^{p-1}(M)$. We find this observation interesting. However 
  it is not clear to us the proper interpretation of this condition within our motivating 
   example.

Also after our paper appeared on the net 
  the different proposal for microscopic description of 
 topological M-theory  has been given in \cite{Anguelova:2005cv}.

\appendix 
 
\Section{Appendix: brackets on $C^\infty (T \oplus \wedge^p T^*)$} 
\label{a:11susy} 

 In this Appendix we collect the relevant properties of the brackets $[[\,\,,\,\,]]$ and
 $[\,\,,\,\,]_c$ defined on the sections of $T\oplus \wedge^p T^*$. The proofs of
 these properties are similar to those presented in \cite{weinstein}, in the context of
 Courant algebroid. 

 On smooth sections of $T\oplus \wedge^p T^*$ we can define the bracket 
\beq
 [[v + \omega, \lambda + s]] = [v,\lambda ] + {\cal L}_v s - {\cal L}_\lambda \omega +
  d(i_\lambda \omega),
\eeq{definvrja}
 which is not skew-symmetric. However it satisfies a kind of Leibniz rule
\beq
 [[A,[[B,C]]\,]] = [[\,[[A,B]], C]] + [[B, [[A, C]]\,]],
\eeq{leibnizrule}
 where $A,B,C \in C^\infty(T \oplus\wedge^p T^*)$.
 The property (\ref{leibnizrule}) is easily proved from the definition (\ref{definvrja}).
 In fact the bracket $[[\,\,,\,\,]]$ makes $C^\infty(T\oplus \wedge^pT^*)$ into a Loday algebra.
  Next we define a new bracket $[\,\,,\,\,]_c$ as anitsymmetrization of $[[\,\,,\,\,]]$
\beq
 [A, B]_c = \frac{1}{2} \left ( [[A, B]] - [[B,A]] \right ).
\eeq{newbralsp}
 The explicite expresion for $[\,\,,\,\,]_c$ is given by
\beq
  [ v + \omega, \lambda + s]_c = [v,\lambda] + {\cal L}_v s - {\cal L}_\lambda \omega
  - \frac{1}{2} d( i_v s - i_\lambda \omega).
\eeq{explfirnap}
 Let us introduce ``pairing'' between two sections of $T\oplus \wedge^p T^*$
\beq
 \langle v + \omega, \lambda + s \rangle = \frac{1}{2} (i_v s + i_\lambda \omega),
\eeq{pairingtwo}
 which is a map 
\beq
(T \oplus \wedge^p T^*) \times  (T \oplus \wedge^p T^*)\,\,\rightarrow \wedge^{p-1} T^*,
\eeq{definalamsdp}
 where $\wedge^0 T^* \equiv \mathbb{R}$.
 Thus the relation between two brackets (\ref{definvrja}) and (\ref{explfirnap}) is as follows
\beq
 [A,B]_c = [[A, B]] - d\langle A, B\rangle.
\eeq{relatbsoa}
 The bracket $[\,\,,\,\,]_c$ does not satisfies the Jacobi identity. However it is interesting 
 to examine how it fails to satisfy the Jacobi identity. Let us introduce a trilinear operator,
 Jacobiator, which measures the failure to satisfy the Jacobi identity
\beq
 Jac(A,B,C)= [\,[A,B]_c,C]_c + [\,[B,C]_c,A]_c + [\,[C,A]_c, B]_c.
\eeq{defjacobiator}
 We can prove the following property
\beq
 Jac(A,B,C) = d\left ( Nij(A,B,C) \right)
\eeq{reljacnei}
 where $Nij$ is the Nijenhuis operator
\beq
 Nij(A,B,C) = \frac{1}{3} \left ( \langle [A, B]_c, C\rangle + \langle [B,C]_c, A\rangle +
 \langle [C, A]_c, B\rangle \right ).
\eeq{nijdefs}
 In order to prove (\ref{reljacnei}) we note that
\beq
 [\,[A, B]_c, C]_c = [[\,[[A, B]], C]] - d\langle [A, B]_c, C \rangle
\eeq{noterpoekal}
 where we have used (\ref{relatbsoa}) and the fact that $[[\omega, C]]=0$
 whenever $\omega$ is closed form. 

As corollary of (\ref{reljacnei}) we can establish a few useful theorems.
 Let us call a subbundle $L \subset T \oplus \wedge^p T^*$  isotropic if 
 for any $A,B \in C^\infty (L)$, $\langle A, B\rangle =0$, where $\langle\,\,,\,\,\rangle$
 is defined by (\ref{pairingtwo}).

\begin{theorem} 
 If subbundle $L \subset T \oplus \wedge^p T^*$ is isotropic and involutive 
 with respect to bracket $[\,\,,\,\,]_c$ then  $Nij|_L=0$ and $Jac|_L=0$.
\end{theorem}

 Thus  the bracket $[\,\,,\,\,]_c$ restricted to isotropic involutive subbundle of 
 $T \oplus \wedge^p T^*$ is a Lie bracket. 
 If we add the requirement of maximality 
 to isotropic condition then there is the following theorem. 
By maximal isotropic subbundle $L$ we mean that if the condition 
$$ \langle v + \omega, \lambda + s\rangle =0$$
 is satisfied for all $(v+\omega) \in C^\infty(L)$ then $(\lambda +s) \in C^\infty(L)$, where
  $\langle\,\,,\,\,\rangle$ is defined by (\ref{pairingtwo}).

\begin{theorem} 
If subbundle $L \subset T \oplus \wedge^p T^*$ is maximally isotropic then the following 
 statements are equivalent:\\
  $\bullet$   $L$ is involutive with respect to $[\,\,,\,\,]_c$\\
 $\bullet$ $Jac|_L=0$\\
 $\bullet$ $Nij|_L=0$
\end{theorem}

 For $p=1$ a maximally isotropic involutive subbundle of $T\oplus T^*$ is called  
 a Dirac structure. Thus for the case $p \geq 2$ we refer to a maximally isotropic 
 involutive subbundle of $T \oplus \wedge^p T^*$ as a generalized Dirac structure.

\Section{Hamiltonian constaints for p-brane}

In this Appendix we remind the elements of Hamiltonian analysis of the standard $p$-brane 
 theory. The $p$-brane theory describes the embedding of a $(p+1)$-dimensional 
  world-volume into a $d$-dimensional manifold $M$. The Nambu-Goto 
   action is given by the volume 
   of the embedded $(p+1)$ manifold
   \beq
    S= - T_p \int\limits_{\Sigma_{p+1}} d^{p+1}\sigma\,\,\sqrt{\det({g_{\mu\nu} \d_a X^\mu \d_b X^\nu})},
   \eeq{nabugotoaction}
 where $g_{\mu\nu}$ is the metric with Euclidean signature on $M$ and $T_p$ is brane tension.
  If we put $T_p=1$ then we choose that $\dim[X]=0$. In order to carry the Hamiltonian 
   analysis we assume $\Sigma_{p+1}= \Sigma_p \times \mathbb{R}$, i.e. 
    $\sigma^a = (\sigma^\alpha, \sigma^0)$ with $\sigma^0$ being the evolution parameter. 
    
Denoting by $p_\mu$ the momenta conjugate to $X^\mu$ and starting from 
the Nambu-Goto action (\ref{nabugotoaction}) the constraints can be worked out 
as \cite{Collins:1976eg}
\ber
&&{\cal H} =  g^{\mu\nu} p_\mu p_\nu - \det(q_{\alpha\beta})\\
&&{\cal H}_\alpha = p_\mu \d_\alpha X^\mu
\eer{codnsyepqap}
where 
\beq
q_{\alpha\beta} = g_{\mu\nu} \d_\alpha X^\mu \d_\beta X^\nu
\eeq{defindal}
 is induced spatial metric on the brane.

\end{document}